\documentclass[prl,amsmath,twocolumn,showpacs,superscriptaddress]{revtex4}

\usepackage{epsfig}

\begin{document}

\title{Infinite invariant densities for anomalous diffusion in optical lattices
and other logarithmic potentials} 
\author{David A.  Kessler}
\author{Eli Barkai}
\affiliation{ Department of Physics, Institute of Nanotechnology and Advanced Materials,  Bar Ilan University, Ramat-Gan
52900, Israel}
%
%
\begin{abstract}

We solve the  Fokker-Planck equation for 
Brownian motion in a logarithmic
potential. When the diffusion constant is below a critical value the
solution approaches a non-normalizable scaling state, reminiscent of 
an infinite invariant density. With this non-normalizable density we 
obtain the phase diagram of anomalous diffusion for this important
 process.
We briefly discuss the consequence for a range of physical
systems including atoms in optical lattices and charges in vicinity 
of long polyelectrolytes.  
Our work explains in what sense the infinite invariant density and
{\em not  Boltzmann's equilibrium}
describes the long time limit of these systems. 

\end{abstract}


\maketitle

 Brownian
particles in contact with a thermal heat bath and
in the presence of a  binding potential field $V(x)$
 attain a  steady state which is the Boltzmann equilibrium distribution 
$W_{steady}(x) =N \exp[- V(x)/k_b T]$. An interesting case
 is the logarithmic potential:
$V(x) \propto \epsilon_0  \ln(x)$ for $x\gg 1$. Inserting
$V(x)$ into the Boltzmann distribution, one finds that
the steady state solution is
described by an asymptotic  power law 
$W_{steady} (x) \sim  N x^{ - \epsilon_0/k_b T}$. 
$\epsilon_0/k_b T$
must be larger than one
 for the normalization $N$ to exist. 
Brownian motion in a logarithmic potential has attracted much attention
since it describes many  physical systems, ranging from
diffusion of momentum of two level atoms in optical lattices \cite{CT,Zoller,Lutz},
single particle models for long ranged interacting systems \cite{Bouchet,Lemou},
the famous problem of Manning condensation describing
a charged particle in the vicinity of a long and uniformly charged
wire (i.e. an ion in vicinity of a long charged polymer)
\cite{Manning}, and very recently  the
motion of nano-particles in an appropriately constructed
 force field \cite{Adam}. 

In this Letter we provide the long sought after 
\cite{CT,Zoller,Lutz,Bouchet,Lemou,Chavanis} 
long time solution of the Fokker-Planck equation
describing the dynamics of  Brownian particles in a logarithmic potential. 
Naively one would expect that in the long time limit the equilibrium
distribution, i.e. the Boltzmann distribution, is reached; however the
logarithmic potential turns out to be much more interesting. 
To start with, 
we point out that  
the second moment  in the steady state
$\langle x^2 \rangle_{steady}$ may diverge, namely 
$\langle x^2 \rangle_{steady} =\infty$ if $1<\epsilon_0/k_b T<3$. 
This behavior is unphysical. In particular
in the context of optical lattices, it 
implies that the averaged kinetic  energy of the atoms is infinite which 
is of course impossible (see details below).
If we view the problem of Brownian diffusion in a logarithmic potential
dynamically, we immediately realize that the process cannot be faster 
than diffusion; namely $\langle x^2 \rangle \le 2 D t$ where 
$D$ is the diffusion constant
so $\langle x^2 \rangle=\infty$ is wrong   
(we will soon derive this bound from the Fokker-Planck equation).
In this sense the steady state solution, e.g.  the Boltzmann distribution,
for a particle in a logarithmic potential, does not describe 
the statistical properties of the problem, even in the limit of long times.

Here we show that  
Brownian particles in a logarithmic potential are characterized
by  an {\em infinite invariant density}. This density is not
normalizable (hence the term infinite); 
however as we show, it does  describe the anomalous behavior of the system.
For example it can be used to obtain correctly the
moments of the process, while the normalizable Boltzmann distribution
completely fails to do so. 
We examine these issues first in the context of diffusion of momenta 
of atoms in an optical lattice, since this system is an excellent
candidate to experimentally test our
predictions in the lab. 
The reader should  note that  our results with some small notational 
 changes describe 
a wide class of Brownian trajectories in the presence 
of a logarithmic potential 
(see discussion below). 

{\em Fokker-Planck equation.} The equation for the probability 
density function (PDF) $W(p,t)$ of the momentum $p$  of an atom  in
an optical trap  is modeled within the semi-classical approximation
according to \cite{CT,Zoller,Lutz}
\begin{equation}
{\partial W \over \partial t} = D {\partial^2 \over \partial p^2 } W - {\partial \over \partial p } F(p) W  .
\label{eq01} 
\end{equation}
The cooling force 
\begin{equation} 
F(p) = - {p\over 1 + p^2} 
\label{eq02} 
\end{equation}
restores the momentum to its minimum while $D$ describes
stochastic momentum fluctuations which lead to heating. 
From the Sisyphus  effect, 
interaction of atoms with the
counter propagating laser beams means that
$D$ is determined by the depth of the optical potential \cite{CT,Zoller,Lutz},
which in turn leads to experimental control of the unusual statistical 
properties of this system \cite{Renzoni}. 
  For
$p\ll 1$ the force is harmonic, $F(p) \sim -p $  
while in the opposite limit, $p\gg 1$, $F(p) \sim - 1/p$. 
The effective potential $V(p) = - \int^p F(p) {\rm d} p= (1/2) \ln(1 + p^2)$ 
is symmetric $V(p)=V(-p)$ and 
$V(p) \sim \ln(p)$
when $p\gg 1$ (we use a dimensionless representation for $p$ \cite{Lutz}). 
The minima of the effective  potential $V(p)$
is at $p=0$ which is of course the ideal 
cooling limit which is not maintained due to the fluctuations
described here by $D$. 

{\em Steady state.} The steady state solution of $W(p,t)$ is found
in the usual way: imposing $\partial W_{steady}/ \partial t=0$ 
from 
Eqs. (\ref{eq01},\ref{eq02}) we have $W_{steady} \propto \exp[- V(p)/D]$.
This solution is normalizable only if $D<1$ and 
for that case one finds the Tsallis distribution \cite{Lutz,Renzoni} 
\begin{equation}
W_{steady}\left( p \right) = 
{\cal N} \left( 1 + p^2\right)^{- 1/( 2 D) }
\label{eq03}
\end{equation}
where ${\cal N} = \Gamma\left( {1 \over 2 D} \right) /[ \sqrt{\pi} \Gamma\left( { 1 - D \over 2 D} \right)]$ is a normalization constant.  
This steady state solution was observed in optical lattice
experiments \cite{Renzoni} where
it was shown that this behavior is tunable, namely one may control $D$ to obtain
different steady state solutions. 
Notice that Eq. (\ref{eq03}) 
exhibits a power law decay for large $p$ which 
is clearly related
to the logarithmic potential under investigation.
From Eq. (\ref{eq03}) we have
\begin{equation} 
\langle p^2 \rangle_{steady} \sim 
\left\{
\begin{array}{l l}
{D \over 1 - 3 D} & 0< D < 1/3 \\
& \\
\infty & 1/3 < D < 1.
\end{array}
\right.
\label{eq04}
\end{equation}
As mentioned in the introduction the behavior $\langle p^2 \rangle_{steady}=\infty$ is unphysical since it implies an averaged kinetic energy which 
is infinite \cite{Katori}.

\begin{figure}[t!]
\centerline{\psfig{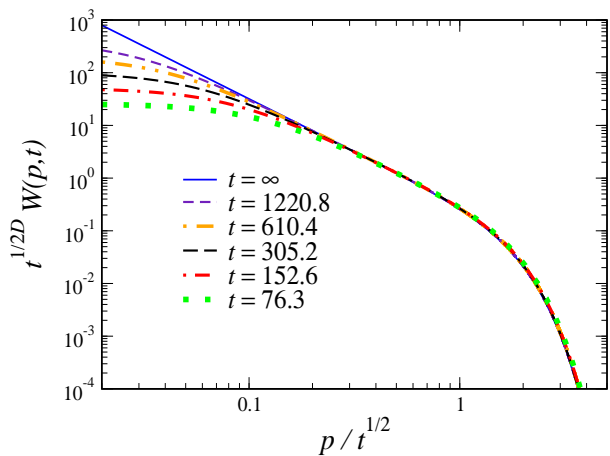}}
\caption{ 
We show that  $f(z)=t^{1/(2 D)} W(p,t)$ $(z=p/t^{1/2})$ 
obtained from numerical solutions of the Fokker-Planck equation 
with a logarithmic 
potential, converges towards the infinite invariant density [solid blue 
$t=\infty$ curve Eq. (\ref{eq12})].
Thus the asymptotic behavior of the system is {\em not} globally
determined by the standard steady state solution. 
 Notice that for $t\to \infty$, $f(z) \sim z^{-1/D}$ for small 
$z$ and hence in in the long time limit the infinite
invariant density  $f(z)$ is non-normalizable. 
Here $D=1/2$ and initially $W(p,t=0)$ is a 
 delta function centered on the origin. 
} 
\label{fig1}
\end{figure}

{\em Bounds on $\langle p^2 \rangle$.} 
To start our analysis we consider the dynamics of $\langle p^2 \rangle$.
Multiplying the Fokker-Planck Eq. (\ref{eq01}) with $p^2$ and integrating over
$p$ we have, after integrating by parts and using the natural boundary condition
that $W(p,t)$ and its derivative at $p \to \pm \infty$ are zero,
\begin{equation}
{\partial \over \partial t} \langle p^2 \rangle = 2 D - 2 \langle {p^2 \over 1 + p^2 } \rangle 
\label{eq05} 
\end{equation} 
where $\langle ... \rangle= \int_{-\infty} ^\infty ... W(p,t) {\rm d} p$. 
Obviously  we have
\begin{equation}
0 \le \langle {p^2 \over 1 + p^2} \rangle \le 1, 
\label{eq06}
\end{equation}
hence $ 2 D -2 \le {\partial \langle p^2 \rangle \over \partial t} \le 2 D$,
and therefore if we start with a finite $\langle p^2 \rangle$ we have 
in the long time limit
\begin{equation} 
(2 D- 2) t \le \langle p^2 \rangle \le 2 D t .
\label{eq07}
\end{equation}
The upper bound clearly implies that $\langle p^2 \rangle$
increases at most linearly  as diffusion persists. 
The lower bound is useful when
$D>1$ since then it shows that $\langle p^2 \rangle \propto t $.
We call the case  $D>1$ the diffusive regime. We now turn to analyze
the cases $D>1$ and $D<1$ separately since they exhibit 
very different behaviors. 

 {\em The case $D<1$.}  We now consider the more interesting case $D<1$ where 
a normalizable  
steady state 
Eq. (\ref{eq03})
exists. In the long time limit
the latter describes well the central part of $W(p,t)$ but not its tails which
govern  the  growth of $\langle p^2 \rangle$ when $1/3<D<1$ 
(for $D<1/3$ higher order moments diverge and the essential problem remains). 
We employ the scaling ansatz
\cite{remarkf}
\begin{equation}
W(p,t) \sim t^{\alpha} f(p/\sqrt{t})
\label{eq08} 
\end{equation} 
which holds for large $p$ and long $t$ and the exponent $\alpha$ will
 be soon determined.  Let us introduce the scaling variable $z= p/t^{1/2}$. This is the typical scaling of Brownian motion, which indicates
that for large $p$ diffusion is in control; however as we now show, 
$f(z)$ in far from a Gaussian so the process is clearly not
simple diffusion. 
Inserting Eq.  
(\ref{eq08}) in the Fokker-Planck equation (\ref{eq01}) 
and using $p\gg 1$ we find
\begin{equation}
D {d^2 f \over d z^2} + \left( {1 \over z} + {z \over 2} \right) {d f \over d z} - \left( \alpha + {1 \over z^2} \right) f = 0. 
\label{eq09}
\end{equation}
For small $z$ we get 
 $f\sim  z^{-1/D}$ or $f\sim  z$;
the latter is rejected since $f(z)$ cannot increase with $z$.
To find $\alpha$ we require that the small $z$ 
solution matches the steady state, since the latter describes well the
density in the center. Using Eq. (\ref{eq08}) with
$f\propto  z^{-1/D}$ we
have
$W(p,t) \propto  t^{\alpha+1/(2 D) } p^{-1/D}$ which is to be matched with the
steady state solution Eq. (\ref{eq03})
 $W_{steady} \propto p^{-1/D}$. Hence we have
\begin{equation}
\alpha=-{1\over 2D}.      
\label{eq10}
\end{equation} 
In this case one  solution of 
Eq. (\ref{eq09}) is immediate: $f(z)= A z^{-1/D}$. While this solution has
the correct small $z$ behavior it does not decay quickly enough at large $z$ \cite{remark0},
so we need the second solution.
The later is found by the method of reduction of order:
%
\begin{equation}
f(z) = A z^{-1/D} \int_z ^\infty s^{1/D} e^{ - s^2 / 4D}{\rm d} s.
\label{eq11} 
\end{equation} 
The constant $A$ is found by matching the small $z$
 solution Eq. (\ref{eq11})
 to the steady
state solution Eq. 
(\ref{eq03}). Solving the integral in Eq. (\ref{eq11})
we reach our first main result
\begin{equation}
f(z) = {{\cal N} z^{-1/D}  \over  \Gamma \left( {1 + D \over 2 D}  \right)}  \Gamma\left( {1 + D \over 2 D }, {z^2 \over 4 D}  \right) 
\label{eq12}
\end{equation} 
where $\Gamma(a,x)=\int_x ^\infty e^{-s} s^{a-1} {\rm d} s$
 is the incomplete Gamma function \cite{Abr}  and
$\Gamma(a)$ is the Gamma function. For small and large $z$ we find
\begin{equation}
f(z) \sim \left\{
\begin{array}{l l}
{\cal N}  z^{-1/D} \ \ &  \ \ z \ll 2 \sqrt{D}  \\
 {  {\cal N} \left( 4 D \right)^{ {1 \over 2} - {1 \over 2 D}} \over \Gamma\left( {1 + D \over 2 D} \right)} z^{-1} e^{- z^2 / 4 D} \ \  & \ \  z \gg 2 \sqrt{D} .
\end{array}
\right.
\label{eq13}
\end{equation} 
Eq. (\ref{eq12}) is non-normalizable since 
according to Eq. (\ref{eq13}) $f(z) \sim z^{-1/D}$ and hence 
$\int_0 ^\infty f(z) {\rm d} z=\infty$.

\begin{figure}[t!]
\centerline{\psfig{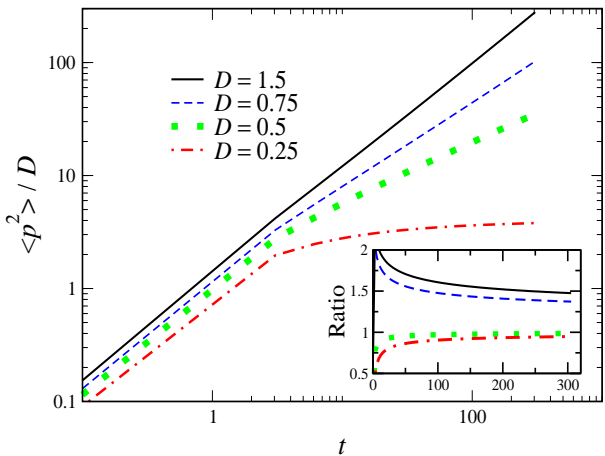}}
\caption{ 
In the long time limit the
 variance $\langle p^2 \rangle$ exhibits three behaviors:
normal diffusion if $D>1$, anomalous diffusion 
$\langle p^2 \rangle\sim t^{3/2-1/(2D)} $ if $ 1/3 < D < 1$  
and steady state behavior $\langle p^2 \rangle \to \mbox{const}$ for
$D<1/3$.  
For very short times (not shown) $\langle p^2 \rangle \sim 2 D t$ since then
the influence of the force field is negligible (initially particles
are on the origin).
In the inset
we show that the ratio of $\langle p^2 \rangle$ obtained from simulations
to $\langle p^2 \rangle$ found from the asymptotic formula Eq. (\ref{eq18})
approaches unity as time increases. 
} 
\label{fig2}
\end{figure}

{\em Infinite invariant density.} We call the  non-normalizable
solution Eq. (\ref{eq12}) an infinite invariant density. 
 In Fig. \ref{fig1} comparison is made between our analytical solution
Eq. (\ref{eq12})  and numerical solutions of
the Fokker-Planck equation. 
As time increases, the solution in the scaled coordinate
 approaches the infinite invariant density
Eq. (\ref{eq12}), 
which describes  the asymptotic scaling solution of the probability density.
For any finite long time $t$, expected
 deviations from the  infinite invariant solution
are found for small values of $z$ (see Fig. \ref{fig1}).
These deviations become negligible at $t \to \infty$ however they are
important since they indicate that the pathological divergence of $f(z)$ 
on the origin is slowly approached but never actually reached, namely
the solution is of course normalizable 
for finite measurement times.

{\em The variance $\langle p^2 \rangle$.} Even though the solution 
Eq. (\ref{eq12})
 is
non-normalizable, it can be used to find the second moment
 $\langle p^2 \rangle$. To see this we introduce a cutoff $p_c$ above 
which our solution Eq. (\ref{eq12}) is valid.
 The variance is calculated
in the usual way exploiting symmetry $W(p,t) = W(-p,t)$
\begin{equation}
\langle p^2 \rangle = 2 \int_0 ^{p_c} p^2 W(p,t) {\rm d} p + 2 \int_{p_c} ^\infty p^2 W(p,t) {\rm d} p. 
\label{eq15} 
\end{equation} 
The first term in Eq. (\ref{eq15}) 
is a constant and can be neglected once the second term
is shown to increase with time. Inserting the infinite
invariant solution Eq. (\ref{eq12})  in the second term of
Eq. (\ref{eq15}) 
\begin{equation}
\langle p^2 \rangle \sim 2 t^{ {3 \over 2} - { 1 \over 2 D} } \int_{p_c/\sqrt{t} }  ^\infty z^2  f(z) {\rm d} z 
\label{eq16} 
\end{equation} 
for $1/3<D<1$. The lower limit in the integral $p_c/t^{1/2}$ goes
to zero when $t \to \infty$ and the diffusion is
 anomalous 
\begin{equation}
\langle p^2 \rangle \sim 2 t^{ { 3 \over 2} - { 1 \over 2 D} }\int_0 ^\infty z^2  f(z) {\rm d} z. 
\label{eq17} 
\end{equation} 
Thus the infinite invariant density 
yields the anomalous diffusion in this model. 
While $f(z) \sim z^{-1/D}$ for small $z$ and is hence non-normalizable,
the integral in Eq. (\ref{eq17}) is finite: the $z^2$ cures the pathology
 of the density at the origin. Solving the integral in Eq. (\ref{eq17}) 
as well as the
diffusive regime soon to be discussed
we obtain 
\begin{equation} 
\langle p^2  \rangle \sim 
\left\{ 
\begin{array}{l l} 
{ D \over 1 - 3 D} & D<{1 \over 3} \\
\  & \\
{16 {\cal N}  \over 2^{1/D}  \Gamma\left( {1 + D \over 2 D} \right) } {D \over 3D - 1}  \left( D t \right)^{ {3 \over 2} - {1 \over 2 D} } \ \  & {1 \over 3}< D < 1 \\
\  & \\
2 \left( D - 1\right) t & 1<D.
\end{array} 
\right.
\label{eq18} 
\end{equation}
For $D<1/3$, $\langle p^2 \rangle$ is time independent and is determined
by  the steady state solution Eq. (\ref{eq03}). 
For the  intermediate regime $1/3 < D < 1$ the diffusion is
anomalous, while for $D$ larger than $1$ it is normal 
in agreement with the simple  bounds we have found, Eq. 
(\ref{eq07}). In Fig. \ref{fig2},  numerical solutions for $\langle p^2 \rangle$ versus time exhibit convergence towards  these types of behavior. 

 A simple argument
for the anomalous diffusion exponent in Eq.  (\ref{eq18}),
 for $1/3 < D<1$,  is found by noticing that
the steady state solution Eq. (\ref{eq03})  describes
the  center part  of the PDF however
 with cutoffs determined by diffusion (i.e.  $|p|<\sqrt{ Dt}$)  
hence
\begin{equation} 
\langle p^2 \rangle \propto \int^{\sqrt{D t}}  _{- \sqrt{D t} } p^2  W_{steady}(p) {\rm d} p \propto 2 \int^{ \sqrt{t}} p^{2 - {1/D} }{\rm d} p \propto t^{{3 \over 2} - {1 \over 2 D } }  .
\label{eq19} 
\end{equation}
To characterize the distribution of $p$ and to find 
$\langle p^2 \rangle$ exactly, 
we need the infinite invariant density which cannot be
obtained by similar simple scaling arguments. 

 {\em The case $D>1$.} We now investigate
the diffusive regime $D>1$ where the
steady state solution is non-normalizable.
  In this case our bound
Eq. (\ref{eq07}) 
yields diffusive behavior $p\propto \sqrt{t}$ which suggests a scaling solution
$W(p,t) = t^{\alpha} f(z)$ where now $\alpha=-1/2$ and
as before $z=p/\sqrt{t}$ \cite{remarkf}. 
Eq. (\ref{eq09}) is still valid  
and its solution when $\alpha=-1/2$
is
\begin{equation}
f(z) =  {z^{-1/D}  e^{ - z^2 \over 4 D} \over \left( 4 D\right)^{1-\beta} \Gamma\left(1 - \beta\right)}
\label{eq20} 
\end{equation} 
where $\beta=(1+D)/(2 D)$. 
Notice that when $D \gg 1$ we have 
 $f(z) \sim \exp( - z^2/ 4 D) /\sqrt{ 4\pi  D t}$ which is expected from
a purely Gaussian diffusive process. Roughly speaking the potential
is responsible for accumulating of particles close to the origin
which yields the $z^{-1/D}$ factor in Eq. (\ref{eq20}). 
Similar to the  
case $D<1$ the solution Eq. (\ref{eq20}) exhibits a divergence on
$z=0$ since $f(z)\sim z^{-1/D}$. However, since now $D>1$ 
the solution is  normalizable
and in this regime we do not find an infinite invariant density.
Finally with Eq. (\ref{eq20}) we obtain $\langle p^2 \rangle$ 
for $D>1$ which was given in Eq. (\ref{eq18}).  

{\em The role of infinite invariant density} in physics
is now briefly discussed. 
 Mathematicians have investigated infinite invariant
measures in the context of ergodic theory for many years \cite{Aaronson}.
 More  recently, 
an infinite invariant density 
\cite{Korabel} 
was shown to describe weak chaos
in the well known intermittent
 Pomeau-Manneville
map (by weak chaos we mean dynamical systems
with zero Lyapunov exponent which still exhibit stochastic behavior). 
 In the context of a model of an electron glass,
the distribution of eigenvalues of a relaxation matrix were found to be
non-normalizable which yields slow relaxations and aging \cite{Amir}. 
Thus three routes to infinite invariant densities 
are: weak chaos, disorder and as we showed here the widely
applicable case of diffusion in a logarithmic potential. 
These vastly different mechanisms 
 all exhibit anomalous diffusion
and the usual ergodic hypothesis breaks down. It therefore 
seems likely and certainly
worthy of further investigation that
a broad range of physical systems which exhibit anomalous kinetics
\cite{Bouchaud,Review,Levybook} 
are described by an infinite invariant densities.

{\em Thermal systems.}
So far we have used the example of the motion of two level atoms in an optical
lattice, a system which is not thermal and its equilibrium distribution
is strictly speaking not a Boltzmann equilibrium. As noted in the introduction
we may consider over-damped Brownian particles
coupled to a thermal  heat bath 
with temperature $T$ and get the same results.  
More precisely consider over-damped Brownian motion in the potential
$\epsilon_0 \ln \left( a^2 + x^2\right)/2$ and diffusion constant
$\overline{D}$ (units $m^2/s$).  With the fluctuation dissipation
theorem \cite{Risken} we have 
\begin{equation} 
{\partial P(\tilde{x},\tilde{t}) \over \partial \tilde{t}}  = { k_b T \over \epsilon_0}  {\partial^2 \over \partial \tilde{x}^2} + {\partial \over \partial \tilde{x}} {\tilde{x} \over 1 + \tilde{x}^2 }   P\left( x,t\right) 
\label{eq21} 
\end{equation} 
which after obvious change of notation is the same as Eq. (\ref{eq01}). 
In Eq. (\ref{eq21})  dimensionless time $\tilde{t} = \epsilon_0 \overline{D} t / a^2 k_b T$ 
and space $\tilde{x} = x/a$ are used.  
More importantly our results are not limited to one dimension.
Indeed an infinite wire of radius $a$,
 with uniform charge density per unit length
 $\lambda$ yields the logarithmic  potential
$V(r) = \lambda ln(r) $ for $r>a$ and $a>0$ must be finite. Such a potential
was considered by Manning \cite{Manning} in the context of ion condensation on 
a long polyelectrolyte. Using cylindrical coordinates it is not difficult
to show that the corresponding three dimensional 
 Fokker Planck equation maps onto a one dimension
problem (for coordinate $r$) which is similar to ours. We do note that 
the limit $a\to 0$ yields new behaviors
 beyond what we discussed here (this is
related to the fact that our potential is finite on $p=0$ while $ln(r) \to -\infty$ for $r \to 0$, i.e. when $a$ approaches zero).  

{\em Summary}. 
Steady state solutions
are commonly  assumed to describe  the long time limit
of dynamics of many thermal and non-thermal systems. 
This assumption is one of the pillars on which statistical
mechanics is built. 
Therefore it was rewarding to find that for the widely applicable process
of Brownian motion in a logarithmic potential, 
an infinite invariant density describes the scaling solution.
Both the well known  steady state solution Eq. (\ref{eq03}) 
and the infinite invariant density Eq. (\ref{eq12}) found here
are needed to characterize the long time solution.
Thus Boltzmann's equilibrium concepts
 while important are clearly not sufficient. 
  
{\bf Acknowledgment} Work is supported by the Israel Science Foundation. EB
thanks E. Lutz and F. Renzoni for useful discussion on the physics of
optical lattices.


\begin{thebibliography}{99}
\bibitem{remark0} If  $f(z) \propto z^{-1/D}$ we have
$W(p,t) \propto p^{-1/D} $
which is  too slow a decay to describe the process
with initial condition centered in the vicinity of the origin.
In fact  $W(p) \propto p^{-1/D}$ yields $\langle p^2 \rangle=\infty$  for 
$1/3  <  D  < 1$ while we proved
$\langle p^2 \rangle \le 2 D t$ so the solution $f(z)\sim z^{-1/D}$ is rejected.
An obvious exception is the case   were  we start the process
in the steady state, 
then one may get $\langle p^2 \rangle=\infty$ for all time which
is valid mathematically but physically wrong.
\end{thebibliography}
\end{document}